# Few-femtosecond phase-sensitive detection of infrared electric fields with a third-order nonlinearity

HANNES KEMPF[1], PHILIPP SULZER[1], ANDREAS LIEHL[1], ALFRED LEITENSTORFER[1], AND RON TENNE[1] *

[1]*Department of Physics and Center for Applied Photonics, University of Konstanz, D-78457 Konstanz, Germany*
*Corresponding author: ron.tenne@uni-konstanz.de*

**Abstract**
Measuring an electric field waveform beyond radio frequencies is often accomplished via a second-order nonlinear interaction with a laser pulse shorter than half of the field's oscillation period. However, synthesizing such a gate pulse is extremely challenging when sampling mid- (MIR) and near- (NIR) infrared transients. Here, we demonstrate an alternative approach: a third-order nonlinear interaction with a relatively long multi-cycle pulse directly retrieves an electric-field transient whose central frequency is 156 THz. A theoretical model, exploring the different nonlinear frequency mixing processes, accurately reproduces our results. Furthermore, we demonstrate a measurement of the real part of a sample's dielectric function, information that is challenging to retrieve in time-resolved spectroscopy and is therefore often overlooked. Our method paves the way towards experimentally simple MIR-to-NIR time-resolved spectroscopy that simultaneously extracts the spectral amplitude and phase information, an important extension of optical pump-probe spectroscopy of, e.g., molecular vibrations and fundamental excitations in condensed-matter physics.

## Introduction

In optical spectroscopy, the frequency-resolved amplitude and phase of light waves encode valuable information about the probed sample. Nonetheless, in time-resolved spectroscopy, a prominent tool for biological, chemical and condensed-matter research, the phase information is not commonly obtained[1,2]. This results mainly from the practical difficulty in measuring phase with linear interferometric techniques over a broad spectral band and a large range of time delays[1,3,4]. Phase information is even of greater significance to resolve the electric field of an optical pulse in the time domain. While for radio-frequency signals an oscilloscope readily performs this task, the equivalent 'optical oscilloscope' still lacks a general solution and is therefore the goal of current research[5,6]. Extending phase-sensitive measurements to the mid infrared (MIR) often presents an additional obstacle, the lack of high-sensitivity and low-noise detectors[7,8].

In the few-terahertz (THz) regime, nonlinear techniques offer an alternative approach that circumvents both of these issues[9,10]. For example, in electro-optic sampling (EOS), nonlinear interaction with a probe pulse, shorter than the half-cycle period of the signal, temporally samples the electric field[11,12]. Moreover, the phase information is transferred from the inconvenient-to-detect THz and multi-THz ranges to the NIR or visible spectral bands where low-noise sensors and arrays are readily available[13,14].

Recently, these considerations motivated a growing interest in the expansion of nonlinear phase-detection methods to the MIR and NIR[5,10,15–19]. This progress supports the ongoing scientific effort in controlling solid-state systems on a few-femtosecond time scale with precisely characterized ultrashort laser pulses[20–24]. Moreover, the addition of the spectral phase information promises to expand the capabilities of existent MIR and NIR spectroscopy and imaging techniques[25]. However, extending EOS to field transients in the MIR and NIR requires the stable generation of probe pulses with a few-femtosecond duration, an extremely challenging experimental task. An alternative successful strategy that circumvents this requirement is to exploit a highly nonlinear interaction between a strong probe pulse and the signal field[5,17,18,26,27]. The high degree of nonlinearity results in a large amplification of the output at the temporal peak of the probe field, effectively creating a gating window shorter than half of the signal's period[17]. Nevertheless, this strategy requires the generation of ultrashort few-cycle probe pulses. The necessity of very stable pulses in conjunction with a high peak field poses an additional experimental hurdle. More recently, a single-shot waveform measurement harnessing third-order nonlinearity has been demonstrated[28]. This implementation relied on µJ-scale pulse energies, typically limiting its application to laser amplifiers with kHz pulse repetition rates. Altogether, a simplified and general scheme to directly measure the spectral amplitude and phase of electric-field transients in the MIR and NIR is an ongoing challenge of high significance.

In this work, we introduce a new method that strives to meet this challenge – multi-cycle third-order sampling (MCTOS). Surprisingly, a phase-sensitive measurement is achieved using third-order nonlinear interactions alone, thus, without generating a sub-cycle gating duration or a carrier-envelope-phase (CEP)-stable signal pulse. The low-order nonlinear signal is realized with sub-nanojoule pulses and can be accurately described with a straightforward perturbative model. Therefore, MCTOS represents a convenient scheme for both pulse characterization and time-resolved phase sensing in the attractive MIR and NIR spectral windows.

## Results and discussion

**Experimental implementation.** In the following, we provide a concise description of the experimental setup; further details can be found in the Methods section. Fig. 1 depicts the schematic setup of MCTOS. The pulsed output of a mode-locked Er:fiber laser oscillator with a 40 MHz repetition rate and center frequency of 193 THz (wavelength of 1550 nm) is separated into two branches to generate the signal and probe pulses. A fiber-coupled electro-optic modulator (EOM) reduces the repetition rate in the signal branch to 20 MHz. Thanks to the strong mode confinement within a highly nonlinear fiber (HNF), the telecom pulse undergoes efficient third-order nonlinear interaction generating spectral components within the range from 130 to 350 THz[29]. This mechanism is applied to synthesize both the signal and the probe pulse in two distinct HNFs. The signal branch features a center frequency of 156 THz (wavelength of 1.92 µm), 1 nJ pulse energy and a duration of roughly 40 fs. The spectrum of the probe covers a full-width at half-maximum bandwidth of $\Delta_{BW} = 61$ THz with a center frequency of $f_c = 246$ THz (corresponding to a wavelength of 1.22 µm) supporting a pulse duration of 12 fs with a pulse energy of 0.2 nJ (probe spectra and pulse characterization are given in Supplementary Note 1).

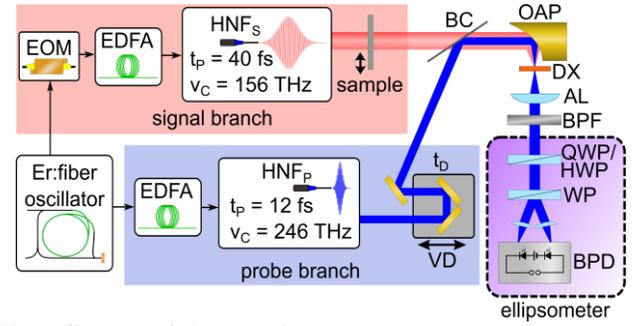

**Fig. 1 Scheme of the experimental setup.** The signal and probe pulses are generated in two distinct highly nonlinear fibers (HNF$_S$ and HNF$_P$). The third-order nonlinear interaction occurs in the detection crystal (DX; Si or GaSe) and the induced polarization change is analyzed by an ellipsometer. EOM: electro-optic modulator. EDFA: erbium-doped fiber amplifier. VD: variable optical delay stage. BC: beam combiner (500 µm thick Si wafer). OAP: off-axis parabolic mirror. AL: achromatic lens. BPF: bandpass filter. QWP/HWP: quarter- or half-wave plate. WP: Wollaston prism. BPD: balanced photodiodes.

A silicon wafer superimposes both beams on an off-axis parabolic mirror (OAP) focusing them into a gallium selenide (GaSe) or silicon (Si) nonlinear detection crystal (DX). The time delay, $t_D$, between the pulses is controlled by a variable delay stage (VD). Both pulses are linearly polarized and orthogonal to each other before entering the DX. The polarization change of the probe is then measured with an ellipsometer whose output is read by a lock-in amplifier (at 20 MHz).

In EOS, the interaction between a signal waveform and a broadband probe pulse within a nonlinear crystal is measured versus the variable time delay between them. Based on second-order nonlinearity, the bandwidth of the EOS probe is required to be as large as the carrier frequency of the signal pulse, 156 THz, supporting a pulse duration of order 3 fs[30]. In addition, EOS requires absolute CEP stability of the signal pulse - a constant phase relation between the envelope of the pulse and the underlying oscillations of the field. None of these conditions are fulfilled in our experiment. Nevertheless, when scanning the time delay between the two pulses, few-fs oscillations emerge (see Fig. 2a). The amplitude of the Fourier transform (FT) of the differential current, shown in the inset, reveals a spectral peak centered at 156 THz, i.e., the carrier frequency of our signal (for the spectral phase, see Supplementary Note 2).

This observation is surprising for several reasons. Not only are the abovementioned conditions for spectral bandwidth violated, but the signal pulses are not phase stabilized and these oscillations are observed even with a silicon DX whose symmetry precludes a second-order nonlinear interaction. In addition to the rapidly oscillating component, a low-frequency constituent appears in the amplitude of the FT. The physical origin of these results is the focus of the following sections.

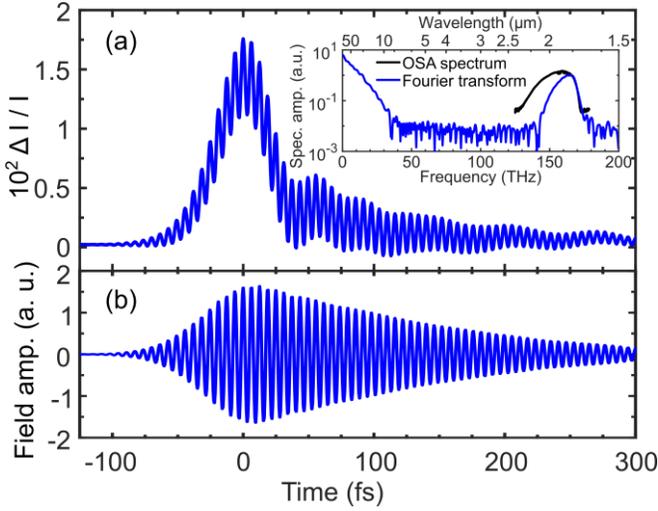

**Fig. 2. Time and frequency domain analysis of the MCTOS output. a** Differential photocurrent as a function of the relative time delay, measured with a 16 μm thick silicon detection crystal in the QWP configuration (see Fig. 1). Inset: Fourier spectrum of the detected signal (blue) plotted semi-logarithmically versus frequency and compared to the spectral amplitude of the signal pulse, as recorded with an optical spectrum analyzer (OSA, black). **b** Field component extracted from the detected signal in **a** by numerically filtering the Fourier transform (FT) (see inset) around 156 THz and applying the inverse FT.

**Intuitive phasor interpretation.** Similar to EOS, the balanced detection scheme employed in MCTOS senses slight variations in the polarization of the probe pulse. These can be described by the generation of new photons through a nonlinear interaction between signal and probe. However, unlike EOS, our results can be explained only by invoking a third-order nonlinearity.

We divide the four-wave mixing interaction into three distinguishable processes that contribute to the MCTOS output. These are depicted in Fig. 3b in a standard arrow scheme; the leftmost arrow is the probe input whereas the fourth arrow from the left stands for the nonlinearly generated probe output. In the first process, referred to as upconversion (UC), a low-frequency probe photon (green arrow) is annihilated and a higher-frequency photon (blue arrow) emerges. Conversely, downconversion (DC) generates an output at a lower frequency. The effective gain and loss of energy in UC and DC, respectively, are represented by black arrows. A third interaction route detected in MCTOS, direct downconversion (DDC), involves two signal photons and was already observed, for example, by Sell *et al.*[31]. In this case, one signal photon is annihilated and another one is created.

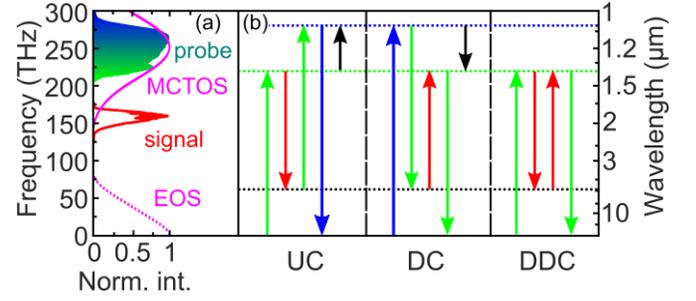

**Fig. 3 Third-order nonlinear processes contributing to MCTOS. a** Normalized spectral intensities of the probe (turquoise color gradient) and signal (red) pulses. The solid and dashed magenta lines represent the spectral response function of the probe pulse for MCTOS and EOS, respectively. **b** Third-order nonlinear interactions. The length and color of the arrows illustrate the energy of the participating photons. The black arrows represent the effective frequency change of the initial probe photon. UC/DC: up-/downconversion. DDC: direct downconversion.

Within a simplified quasi-monochromatic mathematical description, we consider three different angular frequencies for the probe, $\omega$, $\omega_1$, and $\omega + (\Omega - \omega_1)$ along with a single frequency for the signal, $\Omega$. In this case, the complex amplitude of UC and DDC can be written as

$$E_{UC}(\omega, t_D) \propto i\chi^{(3)} A_P(\omega_1) A_P[\omega + (\Omega - \omega_1)]$$
$$\times A_S(\Omega) \exp[i(\Omega t_D + \phi_S - 2\phi_P)], \quad (1)$$

and

$$E_{DDC}(\omega, t_D) \propto i\chi^{(3)} A_P(\omega) \exp[-i\phi_P] A_S(\Omega)^2, \quad (2)$$

respectively. Here, $A_P$ and $A_S$ represent the real spectral amplitudes of probe and signal, respectively. $t_D$ is the adjustable relative time delay between the pulses and $\chi^{(3)}$ is the third-order nonlinear susceptibility. $\phi_S$ and $\phi_P$ are the phases of the signal and probe field, respectively. A full treatment of the broadband case, including all three processes, is given in Supplementary Notes 3-5.

Two significant differences between UC and DDC emerge from these expressions. First, $E_{UC}$ (and $E_{DC}$) oscillates with respect to $t_D$ at the signal frequency whereas the DDC contribution is a constant. This distinction is clear in Fig. 2a where an oscillating term (UC and DC) is offset by a slowly varying transient (DDC). Since the DDC output is clearly distinguishable from the phase-sensitive contribution in the Fourier domain, it can be readily filtered out (see Methods section). An inverse FT of the filtered output reveals the field transient of the signal pulse (Fig. 2b). Second, and more important for MCTOS, only the UC and DC processes depend on the spectral phases $\phi_S$ and $\phi_P$. As shown below, it is thanks to these interaction paths that MCTOS can directly measure the spectral phase of the signal pulse.

Sensing the fields of these four-wave mixing interactions is accomplished through a homodyne detection scheme. Measuring the interference term between the probe and the newly generated photons, the homodyne output can be conveniently described with a phasor expression - the product of the complex amplitudes of the nonlinear output [Eqs. (1)-(2)] and the probe[30]:

$$P_{UC}(t_D) \propto i\chi^{(3)} A_P(\omega) A_P(\omega_1) A_P[\omega + (\Omega - \omega_1)]$$
$$\times A_S(\Omega) \exp[i(\Omega t_D + \phi_S - \phi_P)], \quad (3)$$

$$P_{DC}(t_D) \propto i\chi^{(3)} A_P(\omega) A_P(\omega_1) A_P[\omega + (\omega_1 - \Omega)]$$
$$\times A_S(\Omega) \exp[-i(\Omega t_D + \phi_S - \phi_P)], \quad (4)$$

$$P_{DDC}(t_D) \propto i\chi^{(3)} A_P(\omega) A_P(\omega) A_S(\Omega) A_S(\Omega)$$
$$= i\chi^{(3)} |A_P A_S|^2. \quad (5)$$

Since the polarization of the generated photons is orthogonal to that of the probe, the interference term can be divided into two distinct terms. A field that is in phase with the probe wave results in a rotation of the linear polarization of the probe and corresponds to the real part of the phasor. In contrast, a nonlinear field output with a π/2 phase offset leads to an elliptically polarized interference signal and corresponds to the imaginary portion of the phasor. A balanced ellipsometer equipped with a quarter-wave plate (QWP) detects the imaginary part (π/2-phase-shifted component) of the phasor whereas one with a half-wave plate (HWP) obtains the real part (in-phase component)[30].

Fig. 4 depicts the complex phasors (left) and the corresponding detected waveforms (right) for the three interaction routes described above [Eqs. (3)-(5)]. Since the DDC phasor is not oscillating and purely imaginary (vertical purple bar in Fig. 4a), its output is constant and only measurable with a QWP (Fig. 4b). With an increasing time delay $t_D$, the UC phasor (Fig. 4c, green arrow) rotates anticlockwise in the complex plane with an angular frequency Ω [Eq. (3)] while the DC phasor (Fig. 4c, orange arrow) rotates with the same frequency in the opposite direction. The sum of both phasors (orange-green dashed arrow) constantly points along the imaginary axis, with a periodically varying amplitude. As a result, the total signal from DC and UC is also detectable only in a QWP ellipsometer configuration. Altogether, it is this periodic evolution of the phasor projection that manifests as the phase sensitive MCTOS signal – an oscillation with the carrier frequency of the signal (Fig. 2).

**Comparison with an analytical model.** While the quasi-monochromatic intuitive phasor picture already explains the main features of MCTOS, a rigorous mathematical treatment is required for the case of broadband fields (Supplementary Notes 3-5). In the following, we quantitatively compare the broadband modeling to our results.

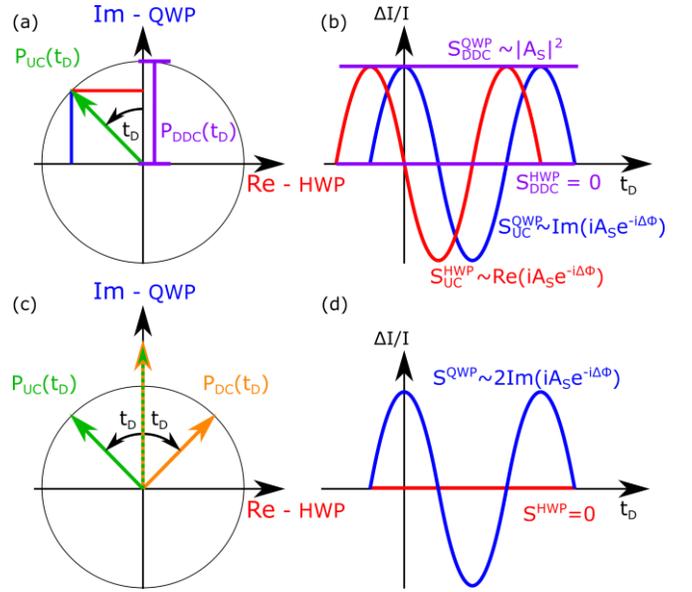

**Fig. 4 Phasor representation of the nonlinear interaction processes.** Upper panels: Upconversion (UC) and direct downconversion (DDC) processes shown separately. **a** Green arrow illustrates the phasor of the UC process for a time delay $t_D$. The blue and red lines indicate the imaginary and real part of the phasor and hence the detected output with a quarter- and half-wave plate (QWP and HWP), respectively. The purple bar represents the phasor of the DDC process. **b** Corresponding UC and DDC outputs depicted versus relative time delay with the same color code as in **a**. Lower panels: The sum of UC and downconversion (DC). **c** Phasors of UC (green arrow) and DC (orange arrow) and their sum (green-orange dashed arrow). **d** Output of the sum of UC and DC when detected with a QWP and HWP in blue and red, respectively. $\Delta\Phi = \Phi_S - \Phi_P$: relative phase difference between signal and probe.

The summed differential current resulting from the UC and DC processes and detected by QWP or HWP configuration can be written as

$$\Delta I^{QWP}(t_D) \propto \int_0^\infty \text{Re}\{[R_{UC}(\Omega) + R_{DC}(\Omega)]$$
$$\times A_S(\Omega) \exp[-i(\Delta\phi^0 + \phi_S^{ho})] \times \exp(-i\Omega t_D)\} d\Omega, \quad (6)$$

$$\Delta I^{HWP}(t_D) \propto \int_0^\infty \text{Im}\{[R_{UC}(\Omega) - R_{DC}(\Omega)]$$
$$\times A_S(\Omega) \exp[-i(\Delta\phi^0 + \phi_S^{ho})] \times \exp(-i\Omega t_D)\} d\Omega. \quad (7)$$

Since both the signal and the probe are derived from the same master oscillator, $\Delta\phi^0$, the CEP difference between them, can be treated as a constant. The higher-order spectral phase of the signal field, given by $\phi_S^{ho} = \phi_S^{(2)} \cdot (\Omega - \Omega_0)^2 + \phi_S^{(3)} \cdot (\Omega - \Omega_0)^3 + ...$, where $\Omega_0$ is the central frequency of the signal. This phase contains the non-trivial information about the temporal form of the electric field. $R_{UC}$ and $R_{DC}$ are the spectral response functions for UC and DC, respectively, which depend both on the spectrum of the probe and the phase-matching of the third-order nonlinear interaction. These expressions confirm the intuitive picture presented in the previous section: The FT of the MCTOS output provides a direct measurement of the spectral amplitude and phase of the signal pulse within the bandwidth offered by the response functions, see Supplementary Note 6. The spectral response functions ($R_{UC}$ and $R_{DC}$) are generally not identical and as a consequence, a phase-sensitive output is expected also for the HWP configuration. In particular, a similar output for the QWP and HWP is obtained when suppressing either the UC or DC contribution.

To experimentally explore this observation, we detect the output with a 25 µm thick GaSe DX and spectrally filter the probe to modify the UC and DC response functions. Essentially, upconverted photons gain energy and therefore tend to appear in the higher-frequency part of the probe spectrum. Inserting 50 nm wide bandpass filters with different center frequencies after the detection crystal (see Fig. 1) spectrally resolves the detection process.

Sections a, c, and e of Fig. 5 depict the output of MCTOS for three different bandpass filters centered at 286, 214 and 250 THz, respectively. The left-hand side shows the detected transients for the QWP (blue) and HWP (red) configurations, while the corresponding amplitudes of their Fourier transforms are presented on the right-hand side.

Applying a BPF centered at the high-frequency edge of the probe (see inset of Fig. 5b), favors upconverted photons (cf. Fig. 3b). Consequently, both ellipsometer configurations (QWP and HWP) measure an oscillating output (Fig. 5a) with nearly identical spectral components (Fig. 5b). This behavior is also predicted by the monochromatic picture (Figs. 4a and b). The numerically calculated spectrum for the UC interaction (green dashed line) demonstrates a high level of agreement with the experimental results. Thus, we are able to suppress the effect of the DC process by spectral filtering. Qualitatively similar results are obtained for a postselection of downconverted photons, achieved by filtering around 214 THz (Figs. 5c, d).

In contrast, isolating the central part of the probe spectrum (around 250 THz) effectively reduces the bandwidth for the detection of the phase-sensitive output. As a result, the oscillating transients detected in MCTOS significantly reduce (Fig. 5e) and their spectra are blue shifted (Fig. 5f). The finite DDC contribution to the output of the HWP configuration (Fig. 5e) is likely due to the dispersion of the detection crystal as further discussed in Supplementary Note 5.

We note that using a Si DX provides similar experimental observations, but their analysis and numerical modeling is less straightforward. The spectral overlap of the Si bandgap with the probe pulse results in a more complex dispersion and absorption spectrum that must be taken into account.

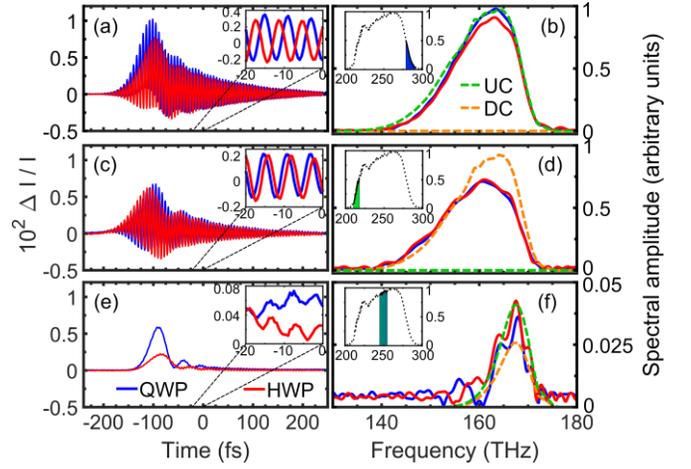

**Fig. 5 Wavelength-resolved analysis of the MCTOS output. a, c, e** Differential current versus time delay measured using quarter- (blue) and half-wave plate (red). The insets show a zoomed-in version of the data. **b, d, f** Corresponding spectral amplitudes with the same color coding as on the left-hand side. The numerical simulations for up- and downconversion are displayed in green and orange dashed lines, respectively. All spectra are normalized to the maximum of the measurement in the QWP configuration in **b**. The insets visualize the normalized probe spectrum (dotted line) with the transmitted spectral range of the bandpass filters (color shaded areas). Measurements were performed with a 25 µm thick GaSe DX.

To discuss the spectral response of MCTOS, we first consider that the homodyne scheme is highly sensitive only within the frequency range of the probe. Thus, detection occurs only if the upconverted or downconverted photon spectrally overlaps with the probe.

Consequently, the sensitivity range of MCTOS is $[f_c - \tfrac{3}{2}\Delta_{BW}, f_c + \tfrac{3}{2}\Delta_{BW}]$ (full magenta line, Fig. 3a), where $f_c$ and $\Delta_{BW}$ are the center frequency and bandwidth of the probe, respectively. In the current setup, this range extends from 155 to 339 THz. The frequency range of MCTOS is thus complementary to EOS whose response vanishes at higher frequencies (dashed magenta line, Fig. 3a).

The bandwidth is noticeable when comparing the spectrum of our first dataset to that obtained with a commercial optical spectrum analyzer (OSA) (black line, inset of Fig. 2a). At frequencies below the sensitivity range of MCTOS, the signal sharply drops with respect to the reference measurement. As the MCTOS transient is simply a Fourier transform of the complex-valued spectrum, such spectral narrowing manifests as temporal stretching of the oscillating component (Fig. 2b) with respect to the original signal pulse.

A second indication for the role of the response function, $R(\Omega)$, is observed of in Fig. 5. Applying a spectral filter at either edge of the probe spectrum broadens the response function of MCTOS in the frequency domain, as thoroughly discussed in Supplementary Note 6. Filtering in detection can also simplify MCTOS in the case of spectral overlap between the signal and the probe. A filter excluding the signal pulse spectrum suppresses the linear interference term between the signal and probe pulses, which may otherwise obscure the MCTOS output. As such, an optimal choice for detection filter rejects the signal spectrum while transmitting only a fraction of the probe in either its high- or low-energy edge. Another factor that impacts the MCTOS response function is the spectral phase of the probe pulse. Naturally, an optimal bandwidth is obtained for a transform-limited probe pulse while significant chirping results in spectral narrowing, as described in detail in Supplementary Note 6. In the current experiment, we synthesize a transform-limited probe pulse, as indicated by the results of a frequency-resolved-optical-gating (FROG) measurement (see Supplementary Note 1). Finally, we note that beyond the bandwidth covered by the current experimental implementation, a significant advantage of MCTOS, in comparison with EOS, is its tunability. The central frequency of the probe branch can be readily tailored to cover a specific spectral region through nonlinear interactions in fiber[29] or free-space[32] optics.

**Application – group index dispersions.** Having confirmed our conceptual and theoretical considerations in the previous sections, we now exploit our quantitative understanding of MCTOS to present its first application: characterizing the phase response of two important optical materials in the MIR. Specifically, we measure the dispersion of the group index of refraction by recording the transients of the signal pulse with and without a specimen.

The group index $n_g(\Omega)$ is readily extracted from the difference of the spectral phases between the two measurements $\Delta\phi(\Omega)$ (see Supplementary Note 1) according to

$$n_g(\Omega) = 1 + \frac{c}{d}\frac{\partial \Delta\phi(\Omega)}{\partial \Omega}, \qquad (8)$$

where c is the speed of light and $d$ the thickness of the specimen.

The extracted dispersion of the group index for germanium (Ge) and gallium antimonide (GaSb) are depicted in black circles in Figs. 6a and b, respectively. The results for Ge show an excellent agreement with published experimental data[33–35] in the entire spectral range covered while providing a much higher spectral resolution. This confirms the phase-sensitivity of MCTOS as well as its high reliability for spectral phase measurement.

For GaSb (Fig. 6b), experimental data is scarce[36] and only two publications[37,38] are available modeling absorption in the spectral region we analyze. While a reasonable agreement is obtained with the sparse experimental data, our measurements strongly deviate from the results of both numerical modeling works.

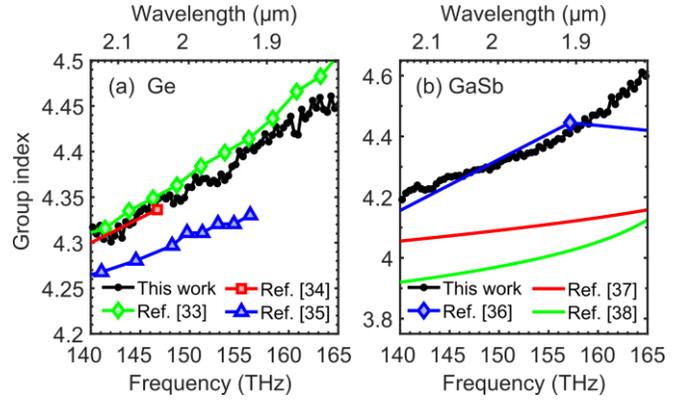

**Fig. 6 Applying MCTOS to obtain the group index of refraction.** Group-index measurements of **a** germanium (Ge) and **b** gallium antimonide (GaSb). Our results (black circles) are compared with several references (colored, see legend). Data points are depicted as markers while the lines are a guide to the eye. Since Refs. [37] and [38] are parameterizations, only a solid line is depicted.

The use of Ge in high-speed electronic components[39] and the potential application of GaSb for next-generation high-electron-mobility transistors (HEMTs)[40] demonstrate the importance of these fundamental semiconductors. Therefore, the sparsity of MIR refractive index measurements, close to the bandgap, is another indication for the lack of a simplified experimental approach for phase-sensitive detection in this spectral region.

## Conclusion

In summary, we present here MCTOS - a new technique for measuring the spectral phase and amplitude of infrared electric-field transients. The pump-probe scheme relying on third-order nonlinear interaction circumvents the requirement for an ultrashort gating pulse altogether. A thorough experimental and theoretical analysis reveals that the detected polarization change of the probe pulses can be understood in the frequency domain as up- and downconversion of near-infrared probe photons similar to the case of EOS. As a proof-of-principle demonstration, we measured the dispersion of the group index of Ge and GaSb.

Unlike conventional EOS suited for low THz frequencies, the sensitivity spectrum of MCTOS is dictated by the center frequency of the probe and can therefore be tailored to fit the spectral content of the signal. Altogether, MCTOS is an excellent candidate to simplify temporally and spatially resolved measurements of the full dielectric function (real and imaginary) in a broad spectral range. As such it may also become valuable for spectroscopic studies of electronic and vibrational transient states, for example, in chemistry and solid-state physics.

## Methods

### Laser source and optical setup

In this work, we employ a custom-built laser system. The laser oscillator is entirely fiber-based and relies on polarization maintaining optical components. Erbium-doped fibers act as an active gain medium and a pigtail saturable absorber mirror (BATOP; SAM-1550-33-2ps) enables stable and self-starting mode locking. A pulse-picking EOM (Jenoptik; AM1550) based on LiNbO$_3$ reduces the repetition rate of the signal branch prior to pulse amplification. The output of the oscillator is split and input to two self-constructed erbium-doped fiber amplifiers (EDFA) that are based solely on fiber components, as well.

Each amplified pulse is coupled into a highly nonlinear fiber, optimized to generate a broadband dispersive wave and a low-frequency soliton for the probe and signal pulse, respectively[41]. For each branch, fine tuning of the output spectra is achieved with a silicon-prism compressor before the pulse enters the HNF. The signal-branch pulse is spectrally filtered with a 150 µm thick gallium antimonide wafer set at Brewster's angle. The filter's second-order dispersion is compensated for with fused silica windows. The probe branch pulse is temporally compressed with a pair of SF10 glass prisms. Inserting a razor blade in the Fourier plane of the prism pair acts as a spectral filter.

The two branches are overlapped in a nonlinear crystal (25 µm thick GaSe or 16 µm thick Si) where third-order nonlinear interactions take place. The GaSe detection crystal is physically exfoliated from a bulk sample, whereas the Si crystal is polished out of a thicker wafer. The output of the nonlinear interaction in the crystal is measured by an ellipsometer setup based on a commercially available balanced photodiodes module (PDB440C, Thorlabs) with two InGaAs detectors. The balanced signal is read out by a radio-frequency lock-in amplifier (UHF, Zurich Instruments) with a 20 MHz reference frequency input.

To filter out the DDC contribution (see Fig. 2), first, a FT of the detected output calculates the spectral amplitude and phase (Fig. 2a). Thanks to their large frequency difference, the DDC contribution (0 to 40 THz) can be separated from the field-sensitive components - UC and DC (140-170 THz). To do so, we multiply the spectral amplitude with a super-Gaussian function, $\exp\left[-\left(\frac{f-f_0}{\Delta f}\right)^6\right]$, with $f_0$ = 155 THz and $\Delta f$ = 80 THz. An inverse FT of the filtered high-frequency component (Fig. 2b) extracts the electric-field transient of the signal.

## Acknowledgements

The work was funded by the Deutsche Forschungsgemeinschaft (DFG) – Project-ID 425217212 – SFB 1432. R. Tenne acknowledges the support of the Minerva foundation.

The authors thank all reviewers for their diligent work and in particular anonymous reviewer number 3 for their helpful suggestions.

## Author contributions

H. Kempf and A. Leitenstorfer posed the original scientific question. H. Kempf designed the experimental setup with support from P. Sulzer and A. Liehl. H. Kempf performed the optical experiments and the simulations. R. Tenne and H. Kempf analyzed the data, refined the physical understanding and prepared the manuscript with significant contributions from all authors.

## Competing interests

The authors declare no competing interests.

## Data availability

Data underlying the results and its analysis script may be obtained from the authors upon reasonable request.

# Supplementary Information: Few-femtosecond phase-sensitive detection of infrared electric fields with a third-order nonlinearity

Hannes Kempf[1], Philipp Sulzer[1], Andreas Liehl[1], Alfred Leitenstorfer[1], and Ron Tenne[1] *

[1]Department of Physics and Center for Applied Photonics, University of Konstanz, D-78457 Konstanz, Germany
*Corresponding author: ron.tenne@uni-konstanz.de

# Table of contents

## Supplementary Note 1: Temporal characterization of the probe pulse

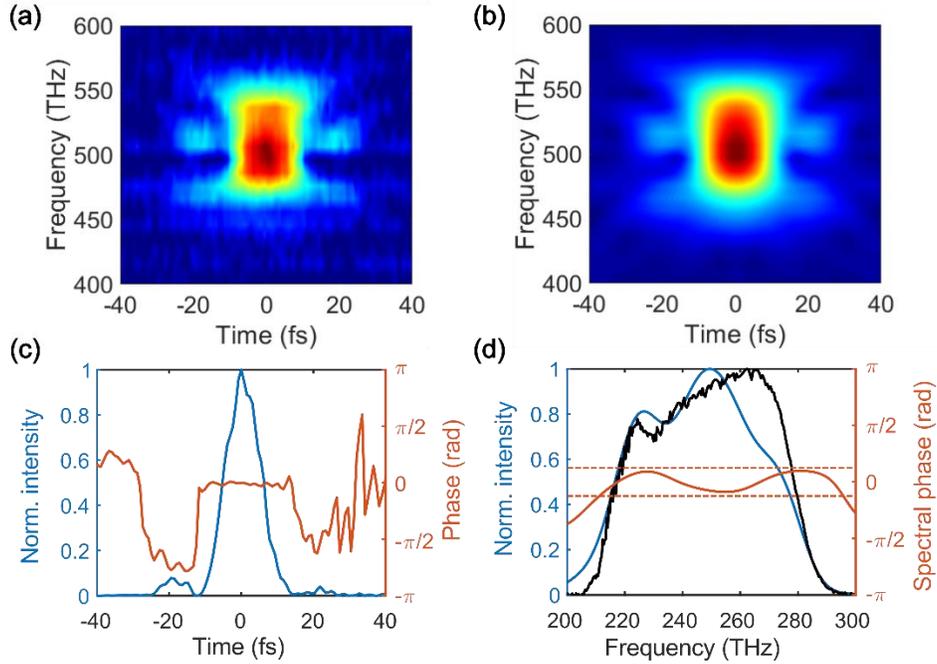

Figure S1. FROG characterization of the probe. **a** and **b** show the measured and reconstructed FROG trace, respectively. **c** The deduced pulse intensity envelope in the time domain. **d** Reconstructed spectrum (blue) and phase (red). Importantly, the phase changes only over a range of $\pm\pi/8$ (dashed red line) over the entire spectrum. For comparison, the spectrum measured with an optical spectrum analyser is shown in black.

In this section, we quantify the spectral phase and pulse duration of the probe pulse. For this purpose, we apply the frequency-resolved-optical-gating (FROG) technique[1] - an extended autocorrelation measurement in which the two replicas of the pulse under investigation are focused into a second-order nonlinear crystal. Sweeping the relative time delay between the two replicas and detecting the spectrum of the sum-frequency generation signal for each time step yields the FROG trace (Fig. S1a). Subsequently, a retrieval algorithm reconstructs the electric-field transient in the time domain to best fit the FROG results. The reconstruction, presented in Fig. S1b, fits well to the measured FROG trace, indicating a successful retrieval. The corresponding pulse in the time domain, shown in blue in Fig. S1c, exhibits a full-width-at-half-maximum (FWHM) duration of 11.4 fs. Its temporal phase (red) is nearly zero during the main pulse. The corresponding spectrum (blue) is depicted in Fig. S1d. Its spectral phase (red) varies in a range of only $\pm\pi/8$ over the entire ~70 THz band exhibiting an almost bandwidth-limited pulse.

To estimate the validity of the measurement and reconstruction algorithm, the spectral intensity analyzed from the FROG measurement is compared the one measured with an optical spectrum analyzer (black line, Fig. S1d). While variations appear, especially in the high-frequency portion, the results are in good agreement. Following this characterization, our modeling for MCTOS in this work (Figs. 5 and 6 of the main text) considers the interaction of a Fourier-limited ultrashort probe pulse with the signal pulse under inspection.

## Supplementary Note 2: Spectrum and phase of a typical MCTOS measurement

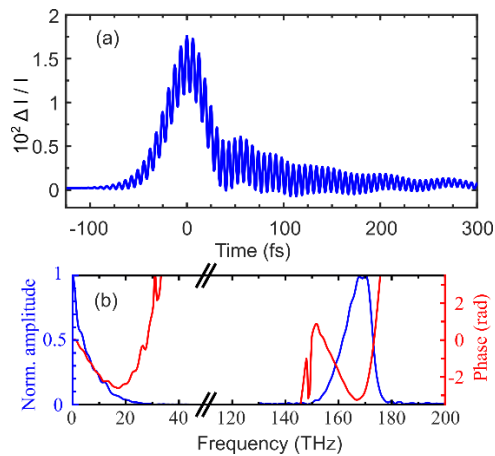

Figure S2. **a** Detected output versus time delay. **b** Fourier transform of **a**.

Figure S2a presents the differential current as in Fig. 2a of the main text. Its Fourier transform is shown in Figure S2b. The spectral amplitude and the corresponding phase are depicted by the blue and red lines, respectively. For better visualization, the amplitudes of both spectral components are normalized individually.

# Supplementary Note 3: Solving the nonlinear wave equation for four-wave mixing

Here, we provide a detailed derivation of the analytical model for all three interactions routes described within the main text. We start by solving the nonlinear wave equation to obtain an expression for the newly generated field, following standard literature, e.g., [2]. In the frequency domain, the expression of the nonlinear wave equation is

$$\frac{d^2}{dz^2}\tilde{E}(\omega,z) + \frac{\omega^2}{c^2} n^2(\omega) \cdot \tilde{E}(\omega,z) = -\frac{\omega^2}{\epsilon_0 c^2} \tilde{P}^{NL}, \tag{S1}$$

where $c$ is the speed of light, $n$ is the refractive index of the detection crystal, $\epsilon_0$ is the vacuum permittivity, and $\tilde{P}^{NL}$ is the nonlinear source term. The electric fields can be expressed as a sum of harmonic components

$$\tilde{E}_n(\omega_n, z) = E_n(\omega_n, z) e^{ik_n z}, \tag{S2}$$

with $E_n(\omega_n, z) \equiv A_n(\omega_n, z) e^{-i\phi_n(\omega_n, z)}$, the wave vector $k_n = n(\omega_n) \cdot \omega_n/c$, and the spectral phase $\phi_n(\omega_n, z)$. For three fields $\tilde{E}_k(\omega_k, z)$, $\tilde{E}_m(\omega_m, z)$, and $\tilde{E}_n^*(\omega_n, z)$ (two photons annihilated and one generated) propagating along the $z$-direction, the nonlinear source term is given by

$$\tilde{P}^{NL}(\omega_p, z) = 4\epsilon_0 \chi_{\text{eff}}^{(3)} \tilde{E}_k(\omega_k, z) \tilde{E}_m(\omega_m, z) \tilde{E}_n^*(\omega_n, z)$$

$$= 4\epsilon_0 \chi_{\text{eff}}^{(3)} E_k E_m E_n^* e^{i(k_k+k_m-k_n)z}, \tag{S3}$$

with the vacuum permittivity $\epsilon_0$. Here, $\omega_p = \omega_k + \omega_m - \omega_n$, is the frequency of the wave component produced by the four-wave mixing interaction. The effective third-order susceptibility $\chi_{\text{eff}}^{(3)}$ is assumed to be real-valued and identical for all processes considered in the following. Inserting these expressions [Eqs. (S2) and (S3)] into the nonlinear wave equation [Eq. (S1)], we obtain a propagation equation for a general frequency component of the electric field $E_p$ that follows the ansatz in Eq. (S2):

$$\left[\frac{d^2 E_p}{dz^2} + 2ik_p \frac{dE_p}{dz} - k_p^2 \cdot E_p + \frac{\omega_p^2}{c^2} n^2(\omega_p) E_p\right] e^{ik_p z}$$

$$= -4\frac{\omega_p^2}{c^2} \chi_{\text{eff}}^{(3)} E_k E_m E_n^* e^{i(k_k+k_m-k_n)z}. \tag{S4}$$

With the definition of the wavevector, we cancel out the third and fourth term on the left-hand side of Equation (S4). Applying the slowly varying amplitude approximation,

$$\left|\frac{d^2 E_p}{dz^2}\right| \ll \left|k_p \frac{dE_p}{dz}\right|, \tag{S5}$$

we can neglect the first term of the left-hand side of Eq. (S4). Introducing the wave-vector mismatch $\Delta k = k_k + k_m - k_n - k_p$ further simplifies Eq. (S4) to

$$\frac{dE_p}{dz} = \frac{2i\omega_p^2 \chi_{eff}^{(3)}}{c^2 k_p} E_k E_m E_n^* e^{i\Delta k \cdot z}. \tag{S6}$$

Under the undepleted-pump approximation ($E_k$, $E_m$, and $E_n$ independent of z), the integration over the thickness $L$ of the nonlinear crystal produces a simple expression for the newly generated field

$$E_p(L) = \frac{2i\omega_p^2 \chi_{\text{eff}}^{(3)}}{c^2 k_p} E_k E_m E_n^* \xi(L), \tag{S7}$$

with the phase-matching factor

$$\xi = \frac{\exp[i\Delta k \cdot L] - 1}{i\Delta k}. \tag{S8}$$

## Supplementary Note 4: Introduction of the three frequency-mixing processes and their complex phasors

After deriving a general expression for the newly generated field through third-order nonlinear interaction, we now introduce the three different processes of MCTOS. For this, we first provide the energy conservation for each of them:

$$\omega_{UC} - (\Omega - \omega_1) = \omega, \tag{S9}$$

$$\omega_{DC} - (\omega_1 - \Omega) = \omega, \tag{S10}$$

$$\omega_{DDC} + (\Omega_1 - \Omega) = \omega. \tag{S11}$$

We consider here the frequencies $\omega_{UC}$, $\omega_{DC}$ and $\omega_{DDC}$ as initial input of the process (even though this definition is somewhat arbitrary) and use $\omega$ to indicate the frequency of the nonlinear source term for all three processes. The frequencies of the photons are arranged according to the order in which they appear in the arrow scheme (Fig. 3b of the main text). For the upconversion (UC) process, for example, the initial probe photon ($\omega_{UC}$; leftmost arrow in green) interacts with a signal photon ($\Omega$; second arrow from the left in red) and another probe photon ($\omega_1$; third arrow from the left in green) to generate a new photon ($\omega$; rightmost arrow in blue).

By separately inserting these frequency relations into Eq. (S7), we obtain the expressions for three nonlinear contributions to a general frequency $\omega$:

$$E_{UC}(\omega) = i \iint_0^\infty \frac{2\omega^2}{c^2 k(\omega)} \chi_{\text{eff}}^{(3)} \xi_{UC} E_P(\omega_1) E_P[\omega + (\Omega - \omega_1)] E_S^*(\Omega) d\omega_1 d\Omega, \tag{S12}$$

$$E_{DC}(\omega) = i \iint_0^\infty \frac{2\omega^2}{c^2 k(\omega)} \chi_{\text{eff}}^{(3)} \xi_{DC} E_P^*(\omega_1) E_P[\omega + (\omega_1 - \Omega)] E_S(\Omega) d\omega_1 d\Omega \tag{S13}$$

and

$$E_{DDC}(\omega) = i \iint_0^\infty \frac{2\omega^2}{c^2 k(\omega)} \chi_{\text{eff}}^{(3)} \xi_{DDC} E_P[\omega - (\Omega_1 - \Omega)] E_S(\Omega_1) E_S^*(\Omega) d\Omega_1 d\Omega, \tag{S14}$$

with the spectral field of the probe $E_P$ and signal $E_S$ pulse. To reduce the number of variables, we replaced the initial frequency of each interaction, invoking the energy-conservation constrains expressed in Eqs. (S9)-(S11). Since we treat broadband spectra, integration over the frequencies $\omega_1$ and $\Omega$ (or $\Omega_1$ and $\Omega$ for DDC) is necessary to obtain all contributions for the newly generated field. The phase-matching factors $\xi_{UC}$, $\xi_{DC}$, and $\xi_{DDC}$ are defined as:

$$\xi_X = \frac{\exp[i \Delta k_X(\omega, \Omega, \omega_1) L] - 1}{i \Delta k_X(\omega, \Omega, \omega_1)}, \tag{S15}$$

where $X$ is a substitute for the three processes UC, DC, and DDC. $L$ is the thickness of the detection crystal and $\Delta k_{UC}$, $\Delta k_{DC}$, and $\Delta k_{DDC}$ are the phase mismatches for each process:

$$\Delta k_{UC} = k[\omega + (\Omega - \omega_1)] + k(\omega_1) - k(\omega) - k(\Omega). \tag{S16}$$

$$\Delta k_{DC} = k[\omega + (\omega_1 - \Omega)] - k(\omega_1) - k(\omega) + k(\Omega). \tag{S17}$$

$$\Delta k_{DDC} = k[\omega - (\Omega_1 - \Omega)] + k(\Omega_1) - k(\omega) - k(\Omega). \tag{S18}$$

We express the spectral fields of the probe and signal pulse as:

$$E_P(\omega) = A_P(\omega) \exp(-i \phi_P^0), \tag{S19}$$

$$E_S(\Omega) = A_S(\Omega) \exp[-i \phi_S(\Omega)] = A_S(\Omega) \exp\{-i[\phi_S^0 + \Omega t_D + \phi_S^{ho}(\Omega)]\}, \tag{S20}$$

with the carrier-envelope phases (CEP) of the probe $\phi_P^0$ and signal $\phi_S^0$, their spectral amplitudes $A_P(\omega)$ and $A_S(\Omega)$, and their relative time delay $t_D$. $\phi_S^{ho}(\Omega) = \phi_S^{(2)} \cdot (\Omega - \Omega_0)^2 + \phi_S^{(3)} \cdot (\Omega - \Omega_0)^3 + \ldots$ denotes the higher-order spectral phase of the signal pulse with $\Omega_0$ being its central frequency. Since the probe pulse is bandwidth-limited, its higher-order spectral phase is zero [3]. Inserting both terms of the spectral fields into Eqs. (S12) - (S14), we obtain:

$$E_{UC}(\omega) = i \iint_0^\infty \frac{2\omega^2}{c^2 k(\omega)} \chi_{\text{eff}}^{(3)} \xi_{UC} A_P(\omega_1) A_P[\omega + (\Omega - \omega_1)]$$
$$\times A_S(\Omega) \exp[i(\Omega t_D + \phi_S^0 - 2\phi_P^0 + \phi_S^{ho})] \, d\omega_1 d\Omega, \tag{S21}$$

$$E_{DC}(\omega) = i \iint_0^\infty \frac{2\omega^2}{c^2 k(\omega)} \chi_{\text{eff}}^{(3)} \xi_{DC} A_P(\omega_1) A_P[\omega + (\omega_1 - \Omega)]$$
$$\times A_S(\Omega) \exp[-i(\Omega t_D + \phi_S^0 + \phi_S^{ho})] \, d\omega_1 d\Omega, \tag{S22}$$

$$E_{DDC}(\omega) = i \iint_0^\infty \frac{2\omega^2}{c^2 k(\omega)} \chi_{\text{eff}}^{(3)} \xi_{DDC} A_P[\omega - (\Omega_1 - \Omega)]$$
$$\times A_S(\Omega_1) A_S(\Omega) \exp\{i[(\Omega - \Omega_1) t_D + \phi_S^{ho}(\Omega) - \phi_S^{ho}(\Omega_1) - \phi_P^0]\} \, d\Omega_1 d\Omega. \tag{S23}$$

The linear polarization of these terms is perpendicular to that of the probe field. The ellipsometer, composed of a waveplate followed by a Wollaston prism and a balanced detector (Fig. 1 of the main text), achieves two effects. First, it mixes the orthogonally polarized fields, producing an interference signal between the probe and the nonlinearly generated field. Second, the ellipsometer balances the intensities on the two diodes so that the difference between them yields the interference term alone.

The interference signal can be divided into an in-phase and a π/2-phase component that form the real and imaginary parts of a complex phasor, respectively. The former results in a rotation of the polarization input into the ellipsometer whereas the latter induces a slight ellipticity. To obtain the expression of the complex phasor, following ref [4], we multiply the nonlinearly generated fields [Eq. (S21)-(S23)] with the complex conjugate of the probe field with the same frequency, $E_P^*(\omega)$. Measuring the spectrally integrated intensity with the photodiodes results in an additional integration over $\omega$:

$$P_{UC}(t_D) = i \iiint_0^\infty \frac{2\omega \omega_c}{c^2 k(\omega)} \chi_{\text{eff}}^{(3)} \xi_{UC} A_P(\omega) A_P(\omega_1) A_P[\omega + (\Omega - \omega_1)]$$
$$\times A_S(\Omega) \exp[i(\Omega t_D + \Delta\phi^0 + \phi_S^{ho})] \, d\omega d\omega_1 d\Omega, \tag{S24}$$

$$P_{DC}(t_D) = i \iiint_0^\infty \frac{2\omega \omega_c}{c^2 k(\omega)} \chi_{\text{eff}}^{(3)} \xi_{DC} A_P(\omega) A_P(\omega_1) A_P[\omega + (\omega_1 - \Omega)]$$
$$\times A_S(\Omega) \exp[-i(\Omega t_D + \Delta\phi^0 + \phi_S^{ho})] \, d\omega d\omega_1 d\Omega, \tag{S25}$$

$$P_{DDC}(t_D) = i \iiint_0^\infty \frac{2\omega \omega_c}{c^2 k(\omega)} \chi_{\text{eff}}^{(3)} \xi_{DDC} A_P(\omega) A_P[\omega - (\Omega_1 - \Omega)]$$
$$\times A_S(\Omega_1) A_S(\Omega) \exp\{i[(\Omega - \Omega_1) t_D + \phi_S^{ho}(\Omega) - \phi_S^{ho}(\Omega_1)]\} \, d\omega d\Omega_1 d\Omega, \tag{S26}$$

with the CEP difference $\Delta\phi^0 = \phi_S^0 - \phi_P^0$, the center frequency of the probe $\omega_c$ and the additional factor $\omega_c/\omega$ considering the fact that the photodiodes detect the photon number and not the intensity. The imaginary part (detected with a QWP) of the phasors represents the difference in ellipticity, while the real part (determined in the HWP case) corresponds to the polarization rotation change.

**Supplementary Note 5: General expressions for the output of the nonlinear interaction routes**

Before explicitly obtaining the results of the detected outputs, we simplify their expressions by defining the spectral response functions $R_{UC}(\Omega)$, $R_{DC}(\Omega)$, and $R_{DDC}(\Omega)$ of the three interactions as:

$$R_{UC}(\Omega) = \iint_0^\infty \frac{\omega}{k^*(\omega)} \xi_{UC}^* A_P(\omega) A_P(\omega_1) A_P[\omega + (\Omega - \omega_1)] d\omega_1 d\omega, \tag{S27}$$

$$R_{DC}(\Omega) = \iint_0^\infty \frac{\omega}{k(\omega)} \xi_{DC} A_P(\omega) A_P(\omega_1) A_P[\omega + (\omega_1 - \Omega)] d\omega_1 d\omega, \tag{S28}$$

$$R_{DDC}(\Omega, \Omega_1) = \int_0^\infty \frac{\omega}{k(\omega)} \xi_{DDC} A_P(\omega) A_P[\omega - (\Omega_1 - \Omega)] d\omega. \tag{S29}$$

The phase-matching factor $\xi_i$ strongly depends on the dispersion of the detection crystal and can significantly modify the response functions. Qualitatively, the dispersion leads to a phase shift between the nonlinearly generated fields at different axial positions inside the crystal and can thus result in destructive interference. In an experiment, to obtain optimal spectral response functions, the influence of the phase-matching factor can be mitigated either by using a thin detection medium [decrease $L$ in Eq. (S15) and (S8)] or by harnessing a birefringent crystal to minimize the phase mismatch [Eqs. (S16) -(S18)].

Considering only UC and DC, the expressions for the measured output are [Eqs. (6) and (7) in the main text]:

$$\Delta I^{QWP}(t_D) \propto \text{Im}[P_{DC}(t_D) + P_{UC}(t_D)] = \text{Im}[P_{DC}(t_D) - P_{UC}^*(t_D)]$$
$$\propto \int_0^\infty \text{Re}\{[R_{DC}(\Omega) + R_{UC}(\Omega)] A_S(\Omega) \exp[-i(\Omega t_D + \Delta\phi^0 + \phi_S^{ho})]\} d\Omega, \tag{S30}$$

$$\Delta I^{HWP}(t_D) \propto \text{Re}[P_{DC}(t_D) + P_{UC}(t_D)] = \text{Im}[P_{DC}(t_D) + P_{UC}^*(t_D)]$$
$$\propto \int_0^\infty \text{Im}\{[R_{DC}(\Omega) - R_{UC}(\Omega)] A_S(\Omega) \exp[-i(\Omega t_D + \Delta\phi^0 + \phi_S^{ho})]\} d\Omega. \tag{S31}$$

To obtain the numerically calculated spectra displayed in Fig. 5 in the main text, we have to slightly modify the spectral response functions. In these experiments, intentionally employing spectral filtering, only a part of the probe spectrum is detected after the interaction. This is implemented by restricting the integration over the probe frequency $\omega$ to a lower and upper limit $\omega_{low}$ and $\omega_{high}$, respectively:

$$\tilde{R}_{UC}(\Omega) = \int_{\omega_{low}}^{\omega_{high}} \int_0^\infty \frac{\omega}{k^*(\omega)} \xi_{UC}^* A_P(\omega) A_P(\omega_1) A_P[\omega + (\Omega - \omega_1)] d\omega_1 d\omega, \tag{S32}$$

$$\tilde{R}_{DC}(\Omega) = \int_{\omega_{low}}^{\omega_{high}} \int_0^\infty \frac{\omega}{k(\omega)} \xi_{DC} A_P(\omega) A_P(\omega_1) A_P[\omega + (\omega_1 - \Omega)] d\omega_1 d\omega. \tag{S33}$$

Inserting these in Eqs. (S30) and (S31) allows us to calculate the detected spectrum for different bandpass filters as displayed in Fig. 5 of the main text.

Finally, we consider the DDC process. The expressions for the detected differential current are obtained by taking the real and imaginary part of the phasor [Eq. (S26)] and substituting the spectral response function [Eq. (S29)]:

$$\Delta I_{DDC}^{QWP}(t_D) \propto \iint_0^\infty \text{Re}\left\{R_{DDC}(\Omega, \Omega_1) A_S(\Omega_1) A_S(\Omega) e^{i[(\Omega-\Omega_1)t_D + \phi_S^{ho}(\Omega) - \phi_S^{ho}(\Omega_1)]}\right\} d\Omega_1 d\Omega, \tag{S34}$$

$$\Delta I_{DDC}^{HWP}(t_D) \propto \iint_0^\infty \mathrm{Im}\left\{R_{DDC}(\Omega,\Omega_1)A_S(\Omega_1)A_S(\Omega)e^{i[(\Omega-\Omega_1)t_D+\phi_S^{ho}(\Omega)-\phi_S^{ho}(\Omega_1)]}\right\}\mathrm{d}\Omega_1\mathrm{d}\Omega. \qquad (S35)$$

Even though these outputs depend on the spectral amplitude and phase of the signal pulse, they do not provide direct access to it. In the intuitive phasor picture given in the main text [Eq. (5)], a finite output is only expected for the detection with a QWP. This conclusion holds as long as dispersion in the detection crystal is neglected, resulting in a purely real response function [Eq. (S29)]. However, since all materials exhibit some dispersion, a reduced but measurable HWP output is present in our experiments.

## Supplementary Note 6: Analysis of the MCTOS spectral response function

In this section, we provide numerical calculations of the spectral response function $R(\Omega)$ for different experimental configurations. These are based on the probe-branch spectrum as measured with an optical spectrum analyzer (see Fig. S1d). We restrict the calculations by setting a flat dispersion for the detection crystal - simplifying the phase mismatch to $\Delta k = 0$ and the phase-matching factor to $\xi = 1$. We note that this approximation is justified when using a thin detection crystal.

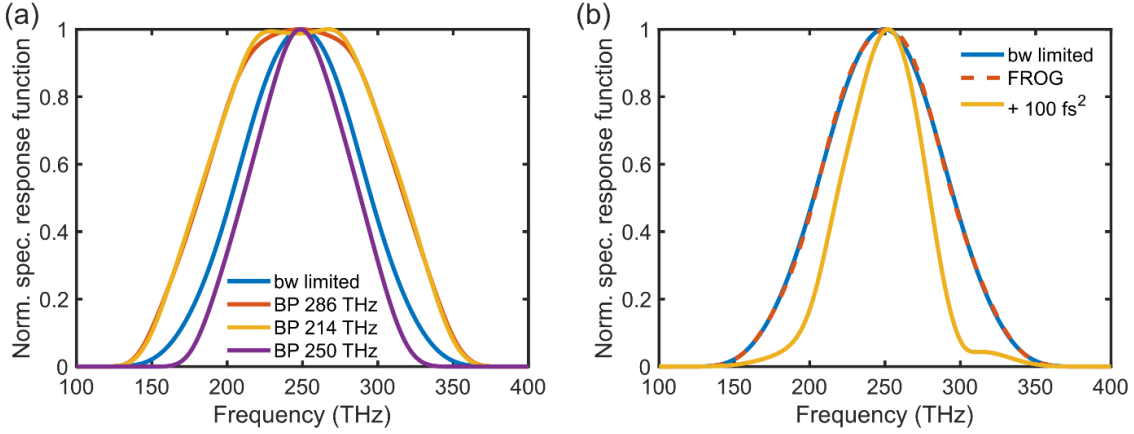

Figure S3. Absolute value of the spectral response functions for different configurations deploying the probe spectrum presented in Fig. S1d (black curve) and a dispersion-free detection crystal. **a** $R(\Omega)$ for a bandwidth-limited probe detected over its entire spectrum (blue) and detected over a narrow band around 286 THz (red), 214 THz (yellow), and 250 THz (purple). **b** The spectral response function assuming a bandwidth-limited probe (blue), the FROG-retrieved spectral phase (dashed red) shown in Fig. S1d, and a linearly chirped probe pulse with a group-delay dispersion of 100 fs$^2$ (yellow).

Fig. S3a analyzes the influence of spectrally filtering the probe after the nonlinear interaction. The absolute value of the sum of the UC and DC spectral response functions $R(\Omega) = R_{UC}(\Omega) + R_{DC}(\Omega)$ is shown for a bandwidth-limited probe considering detection over the full spectral band (blue) and a bandpass filter limiting detection around a center frequency of 284 THz (red), 214 THz (yellow) and 250 THz (purple). For a clear comparison all response functions were normalized to unity at their peaks.

Detecting only at the spectral edges of the probe broadens $R(\Omega)$, whereas a bandpass filter around 250 THz reduces the bandwidth. Since for UC and DC the energy gain (or loss) is large, they are suppressed (see Fig. 3 in the main text) in the latter case. As a result, the effective bandwidth of the probe reduces producing a narrower spectral response function. In contrast, filtering at the edges of the probe suppresses interactions with a low frequency gain or loss and thus increases the bandwidth of $R(\Omega)$.

In addition, we analyze the impact of a chirped probe pulse on the spectral response function (Fig. S3b). For this, the spectral field of the probe has to be extended to including a higher-order phase term $E_P(\omega) = A_P(\omega)\exp\{-i[\phi_P^0 + \phi_P^{ho}(\omega)]\}$. Following the same steps as in sections 3-5, we obtain the spectral response functions:

$$R_{UC}(\Omega) = \iint_0^\infty \frac{\omega}{k^*(\omega)} \xi_{UC}^* A_P(\omega) A_P(\omega_1) A_P[\omega - (\omega_1 - \Omega)]$$
$$\times \exp\{-i[\phi_P^{ho}(\omega) - \phi_P^{ho}(\omega_1) - \phi_P^{ho}(\omega - (\omega_1 - \Omega))]\} d\omega_1 d\omega, \tag{S36}$$

$$R_{DC}(\Omega) = \iint_0^\infty \frac{\omega}{k(\omega)} \xi_{DC} A_P(\omega) A_P(\omega_1) A_P[\omega + (\omega_1 - \Omega)]$$
$$\times \exp\{-i[-\phi_P^{ho}(\omega) - \phi_P^{ho}(\omega_1) + \phi_P^{ho}(\omega - (\omega_1 - \Omega))]\} d\omega_1 d\omega. \tag{S37}$$

Using these equations, we calculate the spectral response functions for the same probe spectrum (black, Fig. S1d), but with different spectral phases. As in Fig. S3a, the bandwidth-limited case is shown in blue. Including the spectral phase retrieved from the FROG measurement (Supplemental 1, Sec. 1) yields only minute changes to the spectral response function (dashed red) validating our approximation of the probe as a bandwidth-limited pulse in all calculations. According to this calculation, adding a linear chirp with a 100 fs$^2$ group-delay dispersion increases the pulse duration of the probe to 21 fs and decreases the bandwidth of the spectral response function (yellow). On an intuitive level, the linear chirp effectively reduces the momentary bandwidth of the probe resulting in a narrower $R(\Omega)$.

**Supplementary Note 7: Derivation of the expression for the group index of refraction**

Here, we derive equation (8) of the main text. The difference of the refractive index $\Delta n(\Omega)$ for a time-domain spectroscopy (TDS) measurement is given by (cf. [5]):

$$\Delta n(\Omega) = \frac{\Delta\phi(\Omega)}{k_0(\Omega)d}, \qquad (S38)$$

with the phase difference $\Delta\phi(\Omega)$ between the measurements with and without sample, the wave vector in vacuum $k_0 = \Omega/c$ and the thickness of the sample $d$. We assume a refractive index in air of 1 and therefore obtain the equation for the refractive index of the specimen $n(\Omega)$:

$$n(\Omega) = 1 + \frac{c}{d}\frac{\Delta\phi(\Omega)}{\Omega}. \qquad (S39)$$

By inserting the previous equation into the definition of the group refractive index[2]

$$n_g(\Omega) = n(\Omega) + \Omega\,\frac{\partial n(\Omega)}{\partial \Omega}, \qquad (S40)$$

we obtain the final expression:

$$n_g(\Omega) = 1 + \frac{c}{d}\frac{\partial\Delta\phi(\Omega)}{\partial\Omega}. \qquad (S41)$$

**Supplementary References**